\documentclass[12pt]{article} 
\usepackage{epsfig,latexsym} 
\usepackage{amsthm}

\input epsf 
\begin{document} 
\font\cmss=cmss10 \font\cmsss=cmss10 at 7pt  
\hfill  NEIP-01-002
 
\hfill LPTENS/01/17

\hfill Bicocca-FT-01-08 \vskip .1in \hfill hep-th/0104026 
 
\hfill 
 
\vspace{20pt} 
 
\begin{center} 
{\Large \textbf{ ${\cal N}=2$ Gauge Theories and Systems \\ with  
Fractional Branes}} 
\end{center} 
 
\vspace{6pt} 
 
\begin{center} 
\textsl{M. Petrini$^{a}$, R. Russo$^{b}$,  and A. 
Zaffaroni$^{c,d}$} \vspace{20pt} 

$^{a}$\textit{Institute de Physique, Universit\'e de Neuchatel, Neuchatel, Switzerland}
  
$^{b}$\textit{Laboratoire de Physique Th\'eorique de l'Ecole Normale Sup\'erieure, Paris, France}

$^{c}$\textit{Universit\`{a} di Milano-Bicocca, Dipartimento di Fisica} 
 
\textit{$^d$ INFN - Sezione di Milano, Italy} 
\end{center} 
 
\vspace{12pt} 
 
\begin{center} 
\textbf{Abstract } 
\end{center} 
 
\vspace{4pt} {\small \noindent 
We analyse the Seiberg Witten curve describing the
${\cal N}=2$ gauge theory dual to the
supergravity solution with fractional branes. 
Emphasis is given to those aspects that are related to stringy
mechanism known as the {\it enhan\c con}.
We also compare our results with the features of the supergravity duals,
which have been variously interpreted in the literature. 
Known aspects of 
the ${\cal N}=2$ gauge theories seem to agree with the  
supergravity solution, 
whenever the two theories can be faithfully compared.} 
\vfill\eject  
\noindent 
\section{Introduction} 

Supergravity solutions dual to non-conformal gauge theories have been
extensively discussed in the past years~\cite{kt}-\cite{mn}. 
%All these backgrounds are dual to strongly coupled gauge theories. 
These backgrounds are typically plagued by IR singularities, but,
for sensible solutions, it is expected that these singularities
can be completely resolved. There are known examples of different
stringy resolution mechanisms.  
The {\it enhan\c con} mechanism~\cite{jpp} is well suited to describe
such resolution for ${\cal N}=2$ models.  
It is the stringy counterpart of well-known effects in the
Seiberg-Witten (SW) solution for ${\cal N}=2$ gauge theories.  
In the ${\cal N}=1$ context, there are examples of deformed conformal
theories which are described via D3-branes expanded into
five-branes by the dielectric effect~\cite{ps}.  
These solutions involve extra brane-like sources. A remarkable example
of a completely regular ${\cal N}=1$ supergravity solution without
extra sources has been constructed by Klebanov and Strassler (KS) in
\cite{ks}.
%It represents the dual of an  
%${\cal N}=1$ gauge theory, realized by wrapping D5-branes on a 2-cycle 
%of a conifold.  Regularity is achieved by deforming the conifold. 
 
In this paper we focus on the physics of four dimensional ${\cal N}=2$
gauge theories in the Coulomb phase. These models naturally arise in
string theory, if one looks at the low energy physics of regular and
fractional D$3$-branes in type IIB theory compactified on
orbifolds like $T_4/Z_2$. The supergravity solutions for regular and
fractional D$3$-branes placed at one of the orbifold singularities
have been extensively discussed in the literature
\cite{kn},~\cite{pg}-\cite{a}.
%We would like to examine 
%if supergravity actually suggests new physics in ${\cal N}=2$ gauge 
%theories.  
In particular, the result for the fractional D$3$-branes %These solutions 
bears some similarities with the ${\cal N}=1$ KS solution, 
namely logarithmic behaviour of the twisted fields and, more suggestive, 
a logarithmic decreasing R-R five-form flux. 
%There are various interpretations in the literature\cite{Pproc,a} 
%of solutions similar to the ${\cal N}=1$ KS solution, 
%with logarithmic behaviour of the warp 
%factor and, more suggestive, 
%a logarithmic decreasing of the five-form R-R flux.   
These features of the supergravity solutions
%Such solutions were 
have been interpreted, on the field theory side, %in the literature,
either as a signal of a possible Seiberg duality in a ${\cal
N}=2$ context~\cite{Pproc}, or as the description of a {\it Higgsed}
vacuum~\cite{a}. 
%At the basis of all ${\cal N}=2$ solutions is the {\it enhan\c con} 
%mechanism~\cite{jpp}. 
It is therefore worthwhile to examine the predictions of the SW solution
for the gauge theories that seem naturally related to these supergravity
solutions. In fact, it is interesting to see whether the main features
of the supergravity solutions are compatible with the known facts on the
field theory side. For the example we are interested in the
SW curve can be constructed following~\cite{witten}. This comparison
%should give strong indications about the behaviour of the supergravity
%solution, 
seems particularly relevant because the type IIB setup with fractional
branes is dual to the M-theory model with $5$-branes, that is used to
determine the SW curve.

Existence of new physics for ${\cal N}=2$ gauge theories is certainly
intriguing.  However, at the level of our analysis, we find that known
aspects of the ${\cal N}=2$ gauge theories seem to agree with the
supergravity solution, whenever the two theories can be faithfully
compared. On the gauge theory side, instanton corrections are
important and responsible for the {\it enhan\c con}.  The study of
such corrections may shed light on how the enhan\c con mechanism is
actually implemented in string theory. In this particular example, the
contribution to the effective action of instantons (corresponding on
the string side to wrapped D1-branes) and higher derivatives terms are
difficult to estimate. Corrections are indeed deeply interconnected
with tensionless string phases of the background. Of course, it would
be interesting to have a better understanding of these corrections.
 
%In this paper, we start an investigation of various ${\cal N}=2$ gauge
%models with attention to some features (and interpretations) of the
%supergravity solutions. 
%Existence of new physics for ${\cal N}=2$
%gauge theories is certainly intriguing.  However, at the level of our
%analysis, we find that known aspects of the ${\cal N}=2$ gauge
%theories seem to agree with the supergravity solution, whenever the
%two theories can be faithfully compared.
 
In Section 2, we briefly review some properties of fractional branes
at an  orbifold singularity.
% and the corresponding supergravity solution. 
In Section~3, 4 and 5, we discuss the SW curve for the associated
${\cal N}=2$ theories.  The weak coupling expansion of the
curve is given, which could be useful for explicit instanton
correction computations. Section~4 is especially devoted to a
discussion of the geometrical aspects of the enhan\c con in these
systems.  We will mainly focus on the origin of 
the moduli space of the ${\cal N}=2$ gauge theory. The gauge theory
enhan\c con is very similar to the original one discussed
in~\cite{jpp}. Much of the novelty comes from the fact that the system
is now defined on a torus in M theory.
In Section~6, we compare the quantum field theory results
with the dual supergravity solutions, which have been discussed in the
literature. In the Appendix, a dictionary for applying AdS/CFT rules
to these systems is given, with attention to few subtleties in the
identification of the moduli/parameters.
  
\section{Fractional branes and ${\cal N}=2$ gauge theories} 

A large class of ${\cal N}=2$ gauge theories can be engineered by
means of physical and fractional D3-branes in orbifold
compactifications of type IIB string theory. We consider the case $R^4/Z_2$ and
choose the coordinates of $R^4$ to be
$(x_6,x_7,x_8,x_9)$.  Particularly important for our construction are
the twisted NS-NS and R-R scalars $(b,c)$. They can be thought as the
flux of the NS-NS and R-R $2$-form along the vanishing $2$-cycle
hidden in the orbifold singularity\footnote{We work with the
convention $2\pi \alpha'=1$}: $2\pi b = \int_{S_2} B$, $2\pi c =
\int_{S_2} C_2$. The perturbative orbifold has value $b=1/2$
\cite{aspinwall}, while $b$ zero or integer correspond to
non-perturbative phases of the theory with tensionless strings. Add
now D3-branes with world-volume $(0123)$ at the point
$x_6=x_7=x_8=x_9=0$. There are two basic types of D3-branes in this
theory: fractional and anti-fractional D3-branes\footnote{The two
types of D3-branes are mutually BPS. We use the name {\it
anti-fractional} with an abuse of language, following the
interpretation as wrapped D5-branes.}. 
Fractional branes have charges
$(b,1/2)$ with respect to the untwisted R-R form $C_{(4)}$ and the
twisted one $C_{(4)}^T$, respectively; anti-fractional branes have
charges $(1-b,-1/2)$.  With a fractional and an anti-fractional
D3-brane we can make a physical D3-brane, whose charge is
$(1,0)$. There are several complementary descriptions for fractional
branes:
\begin{itemize} 
\item{In the perturbative construction of the orbifold, each brane at 
$x_i^{(0)},i=6,7,8,9$ has an image in $-x_i^{(0)}$. A brane and its image 
make up a physical brane, which can be moved at an arbitrary 
point in $R^4/Z_2$. For $x_i^{(0)}=0$, a physical brane appears as a
composite object and can be split in the plane $(x_4,x_5)$.  
The constituents of a physical brane are the two 
types of fractional branes corresponding to the two irreducible
representations of the $Z_2$ action on the Chan-Paton factors. 
%At the orbifold point $b=1/2$, they are  
%interchangeable and have  charge $1/2$ under $C_{(4)}$.
%The orbifold 
%projection acts on the Chan-Paton factors of $n_1$ and $n_2$ 
%fractional branes as 
%\begin{ 
Charges and tensions of these objects
can be determined by the orbifold construction~\cite{dm} or the boundary 
state formalism~\cite{Billo2001}.} 
\item{A fractional brane can be represented as a D5-brane wrapped 
on the collapsed two-cycle of $R^4/Z_2$~\cite{PK3,dMtheory}. 
Similarly an anti-fractional
brane is an anti-D5-brane with a non-trivial gauge field
living on it $\int_{S_2} F=2\pi$~\cite{PK3,dMtheory,Billo2001}.  
This representation is particularly useful when $b\ne 1/2$ and 
the perturbative description of the orbifold is not adequate. 
In general, the induced D3-charges are $b$ and $(1-b)$, while the
tensions are $|b|$ and $|1-b|$. For $b\in [0,1]$, they satisfy the
BPS condition.} 
\item In a useful T-dual picture, the same system is described by 
D4-branes stretched between NS-branes in type IIA \footnote{See, for
example, \cite{lust} for a detailed discussion of the duality between the
Hanany-Witten set-up and the fractional brane systems.}. 
We use standard notations~\cite{witten,hw}. 
The direction $x_6$ is compactified on a circle of radius $L$. 
The two NS-branes have world-volume $(0,1,2,3,4,5)$ and sit at 
$x_6=0$ and $x_6=2\pi bL$, with  $x_7=x_8=x_9=0$.  
The fractional branes can be identified 
with the D4-branes stretched from the first to the second NS-brane,
the anti-fractional branes with the D4-branes stretched from the
second to the first. A fractional and an anti-fractional brane can
join and give a physical D4-brane, which can move away in
$(x_6,x_7,x_8,x_9)$.  
%The orbifold point $b=1/2$ is symmetric under the exchange of the 
%two NS-branes. 
\end{itemize} 
\begin{figure}[h] 
\centerline{\epsfig{figure=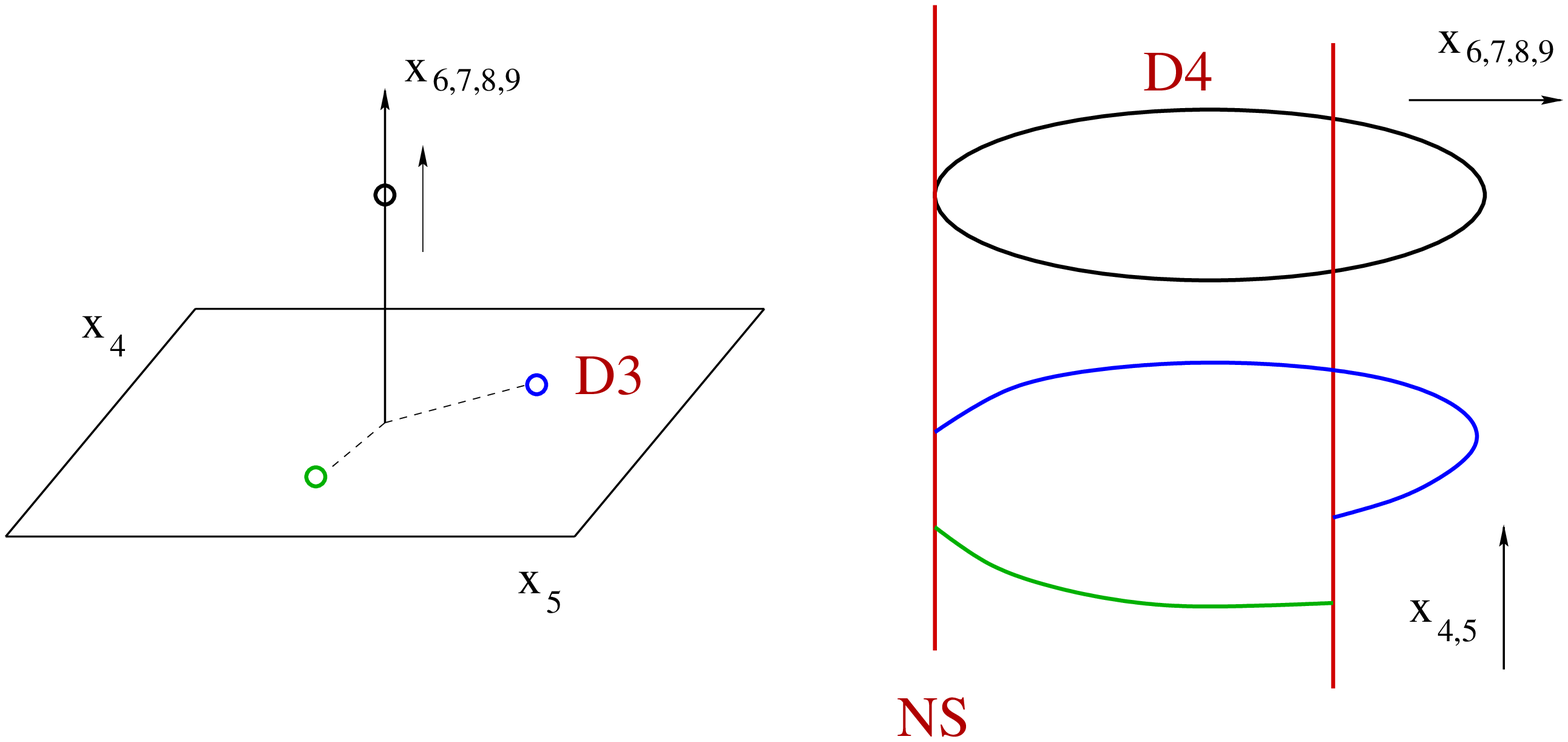,height=5 cm, width=14cm}} 
\caption{Type IIB and IIA picture for physical and fractional branes.} 
\end{figure} 
 
The ${\cal N}=2$ gauge theory living on the D3-branes can be determined 
using the orbifold construction~\cite{dm} or the brane rules~\cite{hw}. 
For $n_1$ fractional  and $n_2$ anti-fractional branes, the gauge theory 
is $U(n_1)\times U(n_2)$ with two bi-fundamental hypermultiplets. 
The gauge couplings of the two groups, $\tau_1,\tau_2$,  
are determined in terms of 
the space-time fields by 
%\begin{equation} 
%\tau_1=b \tau -c,\qquad\qquad \tau_2=(1-b)\tau + c, 
%\label{2} 
%\end{equation} 
\begin{equation} 
\tau_1=(b \tau +c),\qquad\qquad  
\tau_2=(1-b)\tau - c, 
\label{2} 
\end{equation} 
where $\tau=C_0 + i {\rm e}^{-\phi}$ is the complex dilaton of type
IIB. The case $n_1=n_2$ corresponds to a conformal field theory. 
The complex coupling constants of the two groups  
are exactly marginal parameters and the  
theory has an AdS dual: AdS$_5\times$S$_5/Z_2$~\cite{silverstein}. 
When $n_1=N +M$ and $n_2=N$, the theory is no more conformal and the 
coupling constants run at all scales. One of the two gauge factors is not 
asymptotically free and it is ill-defined in the UV.  
All the theories we are interested in can be obtained as suitable  
limits starting from the conformal case. We therefore analyse  
the SW curve for the conformal theory. 
 
\section{The SW curve} 
The curve for the conformal theory $SU(n)\times SU(n)\times U(1)$ with 
two bi-fundamental hypermultiplets was discussed in~\cite{witten}. 
It was obtained by lifting the type IIA configuration with 
NS5 and D4-branes to M theory.  
\begin{figure} 
\centerline{\epsfig{figure=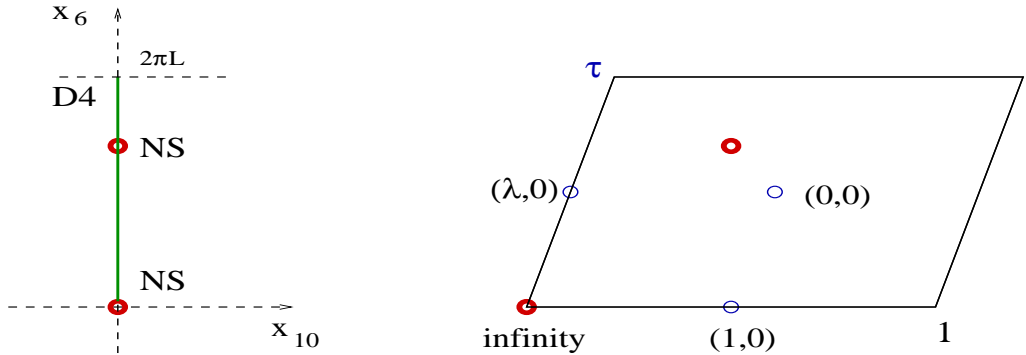,height=5 cm, width=14cm}} 
\caption{The fundamental domain for the torus $E$ and its projection 
to a circle in type IIA with wrapped D4-branes. The four distinguished 
points $x=0,1,\lambda,\infty$ in the cubic representation are 
explicitly indicated.} 
\end{figure} 
  
Call the M-theory coordinate $x_{10}$ ($x_{10}\sim x_{10}+2\pi R$).    
$x_6$ and $x_{10}$ make a  
torus $E$ in M theory, defined by (see Figure 2)  
\begin{eqnarray} 
x_6&\sim& x_6+2\pi L,\nonumber\\ 
x_{10}&\sim& x_{10}+2\pi\theta R. 
\label{c1} 
\end{eqnarray} 
$R$ plays no role in determining the holomorphic data in the SW curve
and it is taken to be large so that we can use the semi-classical
approximation of M theory.
%$\tau=\theta+i/g_s$ can be identified with 
%the type IIB complex dilaton of the previous sections. 
The modular parameter of $E$ can be identified with 
the type IIB complex dilaton of the previous section.
We describe the torus with a cubic equation, 
\begin{equation} 
y^2=x(x-1)(x-\lambda),
\label{c2} 
\end{equation} 
where $\lambda=-\theta_2^4(\tau )/\theta_4^4(\tau )$. 
\begin{figure} 
\centerline{\epsfig{figure=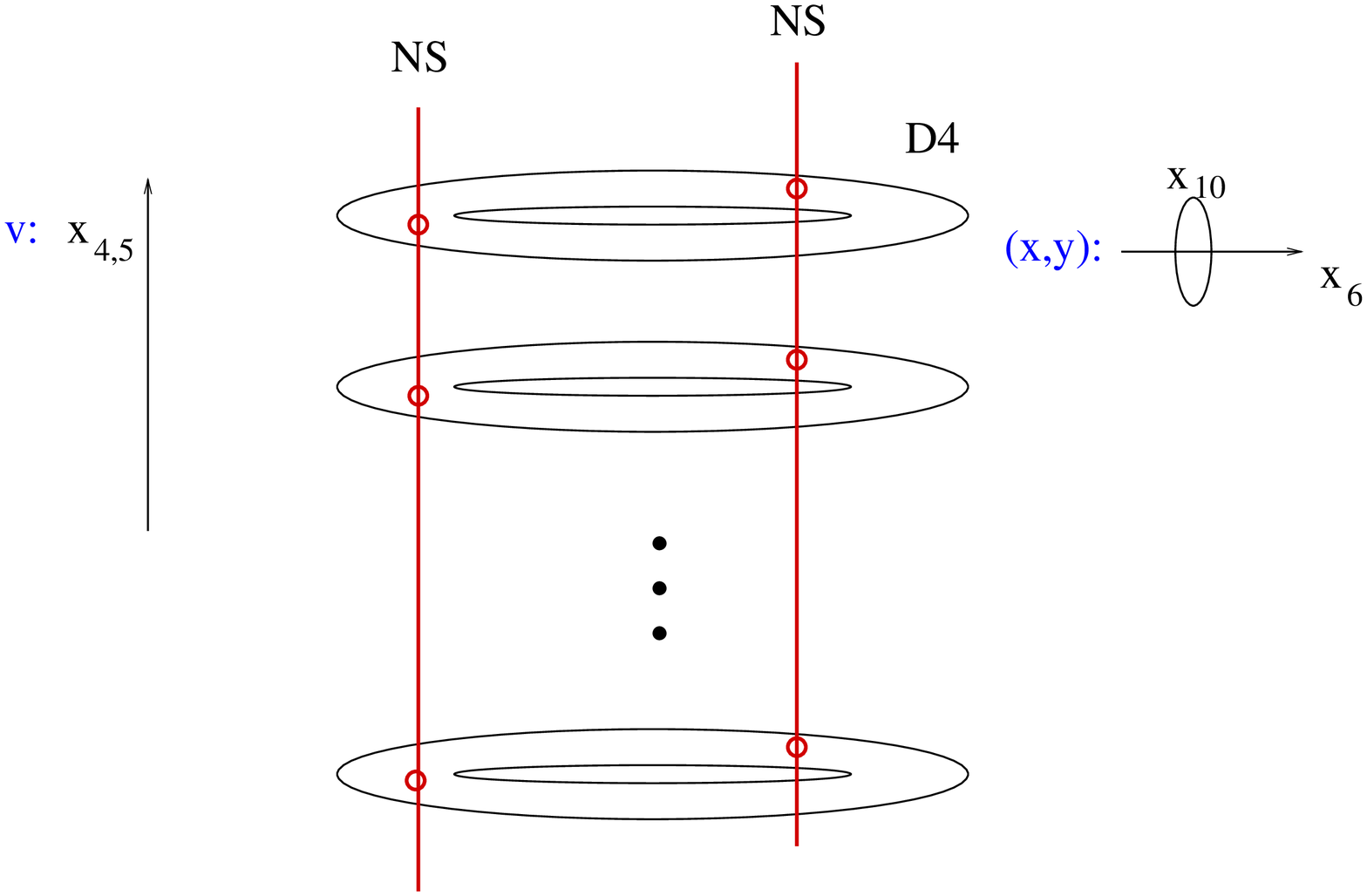,height=7 cm, width=12cm}} 
\caption{The SW curve as $n$-sheeted covering of the torus $E$.} 
\end{figure}  
 
%In the M-theory lift, each D4-brane becomes an M5-brane
%wrapped on the torus $E$. 
The SW curve is an $n$-sheeted covering  
of the torus $E$, $F(x,y,v)=0$, where $F$ is a polynomial of  
degree $n$ in $v$. 
The form of $F$ was determined in~\cite{witten} 
\begin{equation} 
F(x,y,v)=v^n+f_1(x,y)v^{n-1}+...+f_n(x,y).
\label{c3} 
\end{equation} 
Here $f_i$ are meromorphic functions on $E$ with simple poles at the 
positions of the NS-branes. %By Riemann-Roch theorem,  
Given the positions of the poles, each $f_i$ depends 
on two arbitrary parameters.   
For each point $(x,y)\in E$, eq.~(\ref{c3}) gives the positions 
of $n$ branes in the complex plane $v=x_4+ix_5$. $F$ depends 
on $2n$ parameters which represent one mass parameter and the $2n-1$
moduli for the Coulomb branch, 
roughly describing the positions of the fractional branes. 
The generic function $f_i$ is of the form 
\begin{equation} 
f_i(x,y)={ax+by+c\over dx+ey+f},
\label{c4} 
\end{equation} 
with parameters tuned in such a way that  
the two lines $ax+by+c=0$ and $dx+ey+f=0$ have a common intersection on $E$. 
The other two intersections are the two zeros and the two poles of the 
function. The $f_i$'s assume all the complex values twice, with four
double points.   
  
We can choose, for example, 
\begin{equation} 
f_i(x,y)=c_i+d_i{y+y_B\over x-x_B}, 
\label{mero} 
\end{equation} 
which has poles at $P_{\infty}$ and $P_B=(x_B,y_B)$. As discussed in 
\cite{witten}, $c_1$ can be interpreted as the $U(1)$  
modulus and $d_1$ as related to the mass $m$ of the hypermultiplets. 
With some redefinitions, the curve becomes 
\begin{equation} 
\frac{R+S}{2}+\left(\frac{R-S}{2}\right) {y+y_B\over x-x_B}=0~,
\label{curva} 
\end{equation} 
where $R$ and $S$ are polynomial of degree $n$ normalized in such a
way that $R,S=(v^n+...)$. If not explicitly stated, all polynomials in
this paper are normalized so that the higher degree monomial has
coefficient one.  In this representation, the parameters in
$R$ and $S$ are related to the classical positions of the two types of
fractional branes: $R=\prod (v-z_i^{(1)})$ and $S=\prod
(v-z_i^{(2)})$. Indeed for $R=S$ the curve factorizes into $n$ copies
of the torus $E$, describing $n$ physical branes at arbitrary points
(Figure 3). For $R\ne S$, the D4-branes are split and the NS-branes
are bended.
%The explicit factor of $\lambda$ has been introduced  
%for later convenience. 
%We now consider few checks of eq.~(\ref{c6}) that can shed light on 
%the meaning of the various parameters.  
The actual meaning of the polynomials $R$ and $S$ in any corner of 
moduli space is determined by considering the appropriate limit of the
curve.  

For practical computations, it is sometimes convenient to map the
problem to the parallelogram ($u\sim u+1,u\sim u+\tau$,  
see Figure 2). The variable $u$ is immediately  
related to $x_6+ix_{10}$. 
The poles are mapped to $u=0$ ($P_{\infty}$) and $u_B$ ($P_B$). 
%%We will be interested in the  weak coupling limit  
%%$\tau\rightarrow i\infty$ with fixed $u_B$. 
%%In this limit, we can identify $\tau_1=u_B$. 
In terms of the brane construction outlined in the
previous section, we have $u_B=b^{(0)}\tau+c^{(0)}$, where
$b^{(0)}$ and $c^{(0)}$ are the asymptotic values of the twisted fields
$b$, $c$. 
For simplicity, we will focus on the orbifold case
$b^{(0)}=1/2,c^{(0)}=0$, where $\tau_1^{(0)}=\tau_2^{(0)}=\tau/2$.   
Moreover, we move the origin of the $u$-plane away from the position
of the NS-brane by shifting 
%Upon mapping to the parallelogram and 
$u\rightarrow u+\tau/4$. This will simplify the weak coupling
expansion, where it is interesting to focus on one of the two gauge
groups. Following~\cite{schnitzer}, one can rewrite the meromorphic
function $f= \frac{y+y_B}{x-x_B}$ as
\begin{equation} 
f= \frac{\theta_3(u|\tau/2)}{\theta_4(u|\tau/2)} = 
{\theta_3(2u|2\tau) +\theta_2(2u|2\tau )\over \theta_3(2u|2\tau) - 
\theta_2(2u|2\tau )}~.
\label{meme} 
\end{equation} 
The curve then reads 
\begin{equation} 
R\theta_3(2u|2\tau) -S\theta_2(2u|2\tau )=0~. 
\label{paral} 
\end{equation}
The poles are now in $u=\tau/4$ and $u=3\tau/4$. 
The curve is an infinite series in $t=e^{2\pi iu}$. 
As in~\cite{schnitzer}, this infinite-degree polynomial in $t$ 
specifies the positions in the complex $u$ plane
of an infinite number of NS branes. 
%The system represents an infinite repetition of the pattern 
%with two NS branes on the parallelogram.  
We will be interested in the  weak coupling limit
$\tau\rightarrow i\infty$. %with fixed $u_B$.
By truncating to the first order in $q=e^{2\pi i\tau}$ we have, 
\begin{equation} 
-q^{1/4}St+R-q^{1/4}S {1\over t}=0. 
\label{weak} 
\end{equation} 
This is the curve for a gauge group $SU(n)$ with $2n$ flavours 
and coupling constant $\tau_1=\tau/2$~\cite{witten,curve}. 
The moduli are specified by the zeros of the polynomial $R$.
%$R=\prod (v-z_i^{(1)})$ and $S=\prod (v-z_i^{(2)})$. 
The masses for the flavours, which are equal two by two, are given by the 
zeros of the polynomial $S$. This completes the identification of $R$ 
and $S$ in the weak coupling limit.  
At each level of approximation in $q$  
the curve is truncated to a degree $k$ polynomial in $t$ representing $k$ 
NS-branes with $n$ D4-branes stretched between 
them and two sets of $n$ semi-infinite D4-branes on the right and on the 
left. There is a symmetry $t\rightarrow 1/t$ following 
from the symmetric choice of the poles. 
For example, at the next order we have 
\begin{equation} 
qRt^2-q^{1/4}St+R-{q^{1/4}S\over t} +{qR\over t^2}=0. 
\label{weak2} 
\end{equation} 
This is the curve for a $SU(n)^3$ theory, with the first and third factor 
identical.  
 
Non conformal theories can be obtained by considering suitable limits 
in the moduli space. 
Consider, for example, $M$ anti-fractional branes at $z_{\infty}$, 
$M$ fractional branes and $N$ physical branes at $z_i\ll z_{\infty}$
\cite{Pproc,a}.  
We choose, for simplicity, 
a $Z_M$ rotational invariant configuration for the $M$ anti-D3-branes. 
We take therefore 
\begin{equation} 
R=P_{N+M}(v),\qquad\qquad\qquad S=\bar P_N(v)(z_{\infty}^M-v^M)~,
\label{moduli} 
\end{equation} 
with moduli in $P_{N+M},\bar P_N$ much smaller than $z_{\infty}$. 
For $|v|>z_{\infty}$ the theory is conformal. For $|v|<z_{\infty}$, 
the theory reduces at low energies to an $SU(N)\times SU(N+M)$ 
gauge theory with two bi-fundamentals.  
By matching the scales, we define the quantity  
$\Lambda^{2M}=z_{\infty}^{2M}e^{2\pi i\tau_1}=z_{\infty}^{2M}q^{1/2}$ 
appropriate for the IR strongly interacting theory $SU(N+M)$. 
The $SU(N)$ factor is IR free, but its dynamics can be slightly 
modified by the coupling to the other group. 
Since we are interested in the comparison with supergravity, 
we take the t'Hooft limit with $x=Ng_s$ and $y=Mg_s$ fixed. 
In this limit, $\Lambda\sim z_{\infty}e^{-\pi/2 y}$ is kept fixed. 
In order to decouple the cut-off $z_{\infty}$, the appropriate limit  
is $z_{\infty}\rightarrow\infty,y\rightarrow 0$ with $\Lambda$ fixed. 
Since the $SU(N)$ factor is not asymptotically free, the 
$SU(N)\times SU(N+M)$ theory is ill-defined in the UV. For this reason, 
we will keep a finite cut-off in the following.  
In the large $M,N$ limit, the effects of the cut-off manifest 
very sharply near $v\sim z_{\infty}$.  

Given the curve in the form~(\ref{paral}), the computation of the first
instantonic corrections in the weak coupling limit of the 
theory $SU(n)\times SU(n)$ could be explicitly carried out
using the results in~\cite{schnitzer,dhoker}.  

\section{The geometrical picture of the {\it enhan\c con}}

The ${\cal N}=2$ theory has moduli both for the physical and
fractional branes. For every physical brane at $\bar v$, the
polynomials $R$ and $S$ have a common factor $(v-\bar v)$, which 
factorizes in the curve. Factorization of the curve is a signal
of the singularity corresponding to the Higgs branch. 
However, physical branes do not bend the NS branes.
This is the reason why physical branes will
not affect most of our arguments.  

The qualitative behaviour of the coupling constants for the two groups 
is determined by the {\it positions} of the two NS branes on the torus 
as a function of $v$~\cite{witten}. 
These are determined by the solutions of eq.~(\ref{paral})  
for fixed $v$.  
Consider, for simplicity, a configuration with $M$ fractional and $N$ 
physical branes at $v=0$.  
For $|v|<z_{\infty}$ the exact curve~(\ref{paral}) can be expanded 
\begin{equation} 
f(u,\tau)={S+R\over S-R}\sim 1+2\left ({v\over z_{\infty}}\right )^M.
\label{exp} 
\end{equation} 
The two relevant solutions for $u$ can be obtained by expanding $f$  
for small $q$, $f\sim 1+2q^{1/4}(t+1/t)$. Eq.~(\ref{exp}) simplifies to
\begin{equation} 
t+{1\over t}\sim \left({v\over \Lambda}\right )^M. 
\label{exp2} 
\end{equation} 
If this condition is satisfied all higher order terms in $f$ are 
negligible. 

\begin{figure}[h] 
\centerline{\epsfig{figure=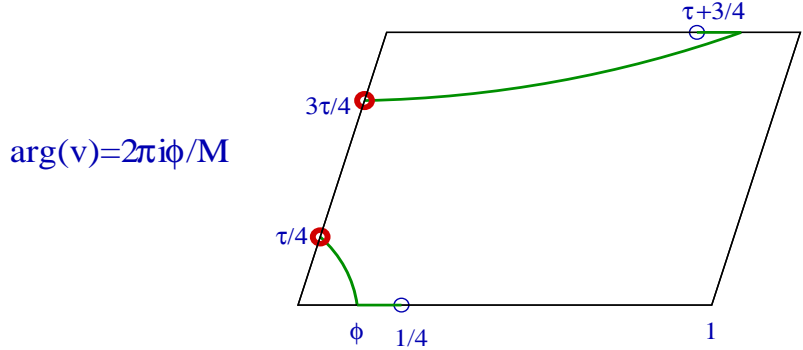,height=5 cm, width=11cm}} 
\caption{Positions of the two NS branes on the torus as functions
of $v$. The argument of the complex number $v$ has been fixed.} 
\end{figure}
Eq.~(\ref{exp2}) is  
the same as the curve for pure $SU(M)$ with scale $\Lambda$. 
The only difference is that the two solutions of~(\ref{exp2}) should 
be brought back to the parallelogram. 
For $|v|>z_{\infty}$ and $|v|<\Lambda$, the positions of the
NS-branes can be directly read from the exact curve~(\ref{paral}),
while for $z_{\infty}> |v|>\Lambda$ they are given by the roots
of the quadratic equation~(\ref{exp2}).
%The position of the two branes are then specified by (see also figure 4) 
At leading order in $M$, one thus obtains (see also figure 4)
\begin{equation}
\begin{array}{cccccccc}
u_1(v)&=\tau/4, \qquad \qquad &u_2(v)=3\tau/4, &|v|>z_{\infty} 
\nonumber\\ 
u_1(v)&=-{M\over 2\pi i}\log v/\Lambda,\qquad &u_2(v)=\tau+{M\over 2\pi i}\log v/\Lambda, &\Lambda<|v|<z_{\infty} 
\nonumber\\ 
u_1(v)&=1/4,\qquad \qquad  &u_2(v)=\tau+3/4, &|v|<\Lambda. 
\end{array}
\label{positions} 
\end{equation} 
As usual in the large $M$ limit, corrections to these formulae rise up 
very sharply near $z_{\infty}$ and $\Lambda$, and make the previous 
expressions completely smooth.  
 
From the type IIA picture, we can roughly estimate the coupling constants 
for the two groups as  $\tau_1(v)=u_1(v)-u_2(v)+\tau$ 
and $\tau_2(v)=u_2(v)-u_1(v)$. 
Note that $\tau_1+\tau_2=\tau$ is then automatically valid. 
We conclude that
\begin{equation} 
\tau_1(v)= {-2M\over 2\pi i}\log v/\Lambda,\qquad \tau_2(v)=\tau+{2M\over 2\pi i}\log v/\Lambda,\qquad\qquad \Lambda<|v|<z_{\infty}. 
\label{1-loop} 
\end{equation} 
This result agrees with the one-loop beta-function for the
$SU(N)\times SU(N+M)$ theory. In the region $\Lambda<|v|<z_{\infty}$
instantonic corrections are indeed suppressed in the large $N,M$
limit.  
They show up for $|v|\leq \Lambda$ where they force the branes in
the positions $v=1/4$ and $v=3/4$. The previous qualitative argument
suggests that, in the large $M$ limit, $\tau_1(v)$ stops running at
$\Lambda$: $\tau_1(v)=1/2$ and $\tau_2(v)=\tau-1/2$ for
$|v|<\Lambda$. A negative imaginary part for $\tau_1$, as suggested by
the perturbative result~(\ref{1-loop}), would be unphysical.
 
Notice that, depending on the phase of $v$, the logarithmic behaviour
of (\ref{positions}) seems to suggest that the NS branes can touch at
$|v|=\Lambda$. Actually, they touch at exactly $2M$ points
$v\sim\Lambda e^{2\pi ik/2M}$, $k=0,1,..,2M-1$, close to a the circle
of radius $\Lambda$. These are the branch-points of the approximate
curve~(\ref{exp2}) and they do not signal any singularity, since the
curve is completely regular there.  They give a reliable picture of
the enhan\c con mechanism, as in the original example
\cite{jpp}.  When the ${\cal N}=2$ system is realized with branes,
even if all the branes are {\it classically} at the origin (the {\it
classical} QFT VEV's are zero), in the quantum theory they are
disposed on a circle of radius $\Lambda$.  If we send in another
fractional brane, the classical value $v$ covers the complex plane,
but quantum mechanically the brane dissolves in the enhan\c con when
$v\sim \Lambda$ and never enters the region $|v|<\Lambda$~\cite{jpp}.
This is easily seen by considering the curve for $M-1$ fractional
branes classically at the origin and one at $v=\phi$:
%$P_{N+M}=v^{N}(v+\phi/(M-1))^{M-1}(v-\phi)$. 
$P_{N+M}=v^{N+M-1}(v-\phi)$.  As in~\cite{jpp}, for
$|\phi|>\Lambda$ there are $2M-2$ branch-points close to the circle
$|v|=\Lambda$ and two at $v\sim\phi$, corresponding to a brane moving
outside the original enhan\c con.  For $|\phi|<\Lambda$ there are $2M$
branch-points close to the circle $|v|=\Lambda$, corresponding to the brane
dissolved in the enhan\c con.

This discussion focused on the low-energy regime and it is appropriate
to describe the group $SU(N+M)$ living on the fractional branes.
Anti-fractional branes can be included in this geometrical picture of
the enhan\c con by considering the exact curve~(\ref{paral}). The
meromorphic function~(\ref{meme}) has double points at
$u=0,1/2,\tau/2,(\tau+1)/2$.  These points are the analogue for the
exact curve of the coincident solutions for $t$ in the
approximate equation~(\ref{exp2}). Thus they can be interpreted as points
where the NS-branes touch. Focusing on the double points on the torus,
in the $v$-plane one can find the values of the branch points.  The
NS-branes touch for exactly $4M$ values of $v$, which have to be
related to a total of $M$ fractional and $M$ anti-fractional branes.
The branch-points at $u=0,1/2$
\begin{equation}
{R+S\over S-R}=f(0),f(1/2)
\qquad\rightarrow\qquad v\sim\Lambda e^{2\pi ik/2M},\, k=0,...,2M-1,
\label{oldbranch}
\end{equation}
have been already discussed and
correspond to fractional branes disposed on a circle of radius $\Lambda$.
It is possible to verify that the new branch-points at $u=\tau/2,(\tau+1)/2$
determine the $2M$ values
\begin{equation}
{R+S\over S-R}=f\left({\tau\over 2}\right),f\left({\tau+1\over 2}\right)
\qquad\rightarrow\qquad v\sim z_{\infty}e^{2\pi ik/2M},\, k=0,...,2M-1~,
\label{newbranch}
\end{equation}
corresponding to the anti-fractional branes disposed at 
circle at $z_{\infty}$.
We can also verify that, if we send in an anti-fractional brane, 
it can move freely in the region $|v|<z_{\infty}$, as expected for
a constituent of an IR free gauge group. If we replace $S\rightarrow
v^N(v^{M-1}-z_{\infty}^{M-1})(v-\bar\phi)$ in the previous formulae
we find from eq.~(\ref{newbranch}) $2M-2$ branch-points at 
$|v|=z_{\infty}$ and two at $|v|\sim\bar\phi$, for every value $\bar\phi<
z_{\infty}$. This corresponds to an anti-fractional probe that is
free to move everywhere, even below the enhan\c con scale.
What is amusing is that eq.~(\ref{oldbranch}) predicts 
in addition two extra branch-points at $|v|\sim\bar\phi$ if 
$\bar\phi<\Lambda$. It looks like a fractional brane has been
unchained and follows the anti-fractional branes below the enhan\c
con. This has a natural interpretation: If we start moving the
cut-off branes below the scale $\Lambda$, the enhan\c con should be
deformed and gradually disappear. Indeed,
without cut-off there would be no asymptotically free group 
at low-energies and thus the enhan\c con is not needed!

We obtained a simple geometrical picture of the enhan\c con mechanism
by taking sections of the curve at various symmetric points on the
torus. It is clear that this interpretation can not be literally
translated in terms of the D-brane setup, since close to the 
enhan{\c c}on locus the geometry is subtle~\cite{johnson}. 
However, the field theory analysis suggests that at the scale
$\Lambda$ the dynamics of all kind of the D-branes is heavily affected
by instanton corrections.

\section{A different limit} 
In this Section, 
we consider a different scaling limit, slightly outside the  
purposes of our paper. This may serve as a consistency check of the 
discussed SW curve.  
In the limit where the coupling constant of one of the two gauge
factors is finite, while the other goes to zero ($\tau_2\rightarrow
i\infty$, $\tau_1$ fixed), we should recover the curve for the conformal 
$SU(n)$ theory with $2n$ flavours.  
In the type IIA picture, 
it can be accomplished by sending the radius $L$ of the $x_6$ circle 
to infinity. We have two NS-branes with $n$ D4-branes stretched 
in between, and two sets of $n$ semi-infinite D4-branes on the 
right and on the left at the same position in $v=x_4+ix_5$. 
This limit requires $b\rightarrow 0$, thus it is natural to expect a
breakdown of the supergravity approximation. 
%$\tau_1=b\tau+c$ remains finite in this limit. 
The limit is conveniently studied using a different representation 
of the meromorphic function than eq. (\ref{mero}). 
The torus~(\ref{c2}) has a $Z_2$  automorphism $y\rightarrow -y$ with 
four fixed points at the values $x=0,1,\lambda,\infty$. 
On the natural variable for the fundamental domain pictured in Figure 2, 
the automorphism acts as $u\rightarrow -u$. Without loss of
generality, we can choose the poles of the meromorphic function to be
$(y_B,x_B)$ and $(-y_B,x_B)$ so that the NS-branes are in 
a $Z_2$ invariant position
\begin{equation} f_i=a_i+\lambda {b_i\over x-x_B}. 
\label{c5} 
\end{equation} 
All the points on $E$ where 
$f_i$ assumes the same complex value are then paired by $Z_2$. In
particular, $x=0,1,\lambda,\infty$ are the four double points. 
With obvious redefinitions, the curve can be written as
\begin{equation} 
(T+V)+\lambda {(T-V)\over x-x_B}=0, 
\label{c6} 
\end{equation} 
where $T$ and $V$ are polynomials of degree $n$. 
Take $\tau\rightarrow i\infty$ (which corresponds to 
$\lambda\rightarrow 0$) and 
rescale $x\rightarrow \lambda x, y\rightarrow \lambda y$ 
with $x_B=\lambda\hat x_B$. We have
\begin{eqnarray} 
y^2&=&-\left (x-{1\over 2}\right )^2+{1\over 4},\nonumber\\ 
(T+V)&=&{(T-V)\over x-\hat x_B}. 
\label{c7} 
\end{eqnarray} 
By eliminating $x$, we obtain the known curve for $SU(n)$ with $2n$  
flavours~\cite{witten,curve}, 
\begin{equation} 
t^2 = P^2 + f^2 Q^2,  
\label{c8} 
\end{equation} 
where  
\begin{eqnarray} 
Q&=&{T+V\over 2},\nonumber\\ 
P&=&{(\hat x_B+1/2)V+(\hat x_B-3/2)T\over 2\hat x_B-1},\nonumber\\ 
f^2&=&-(2\hat x_B-1)^{-2},\nonumber\\ 
t&=&f(R+S)y.
\label{c9} 
\end{eqnarray} 
%  
%It is the same curve discussed in Section 6 with ${\bar t}=t+P$. 
Here $P=\prod_i^n(v-u_i)$ and $Q=\prod_i^n(v-m_i)$ are the degree $n$  
polynomials determining the $SU(n)$ moduli and the masses of the 
flavours, respectively.  
The SW differential $v\,dx/y$ on $E$ indeed has poles at the masses $m_i$ 
with residues proportional to $m_i$, while
$f$ determines the surviving coupling constant~\cite{witten,curve},
since, at weak coupling, $f\sim e^{\pi i\tau_1}\rightarrow 0$. 
 
The CFT $SU(n)$ with $2n$ flavours is not easily
obtained in the AdS/CFT correspondence, using supergravity only. 
One obvious problem is the
large global symmetry. Another point signalling the breakdown of the
supergravity approximation is related to the form of the conformal
anomaly, which is usually written in terms of two coefficients $a$ and
$c$. In the theory under consideration $a$ and $c$ are not equal
already at leading order in $n$. Thus supergravity,
which always requires $a$=$c$, is not enough to describe 
such CFT.
The type IIB orbifold with $b\rightarrow0$ we are using
is indeed a {\it stringy} background.
%It would be interesting to study the consistency of this scaling limit. 
On the other hand, we notice that quantum field theory instantons are
mapped to D1 instantons in the string background. The contribution of
D1 instantons survives in the scaling limit where $\tau_1$ is kept
fixed.  The instanton moduli space of the theory $SU(n)$ with $2n$
flavours in the large $n$ limit was studied in~\cite{mattis}.  The
result AdS$_5\times$S$^1$ is appropriate for D1-instantons, which are
localized on the fixed plane.

\section{Comparison with the supergravity solution and discussion} 
The supergravity solution corresponding to ${\cal N}=2$ fractional branes 
has been extensively discussed in~\cite{kn,bertolini,Pproc,pg}.  
We consider all the branes
in $x_6=x_7=x_8=x_9=0$ and arbitrarily distributed in the $(x_4,x_5)$  
plane. It is convenient to introduce the complex variable  $z=x_4+ix_5$ and to denote the positions of the fractional and anti-fractional branes by  
$z^{(1)},z^{(2)}$, respectively. 
In the gauge theory these correspond to  
VEV's of the Cartan 
values of the adjoint scalars parameterizing the generic vacuum.  
 
Following~\cite{kn,g} we define\footnote{Our conventions are slightly
different from the ones of these papers. Here $(b,c)$ have
periods normalized to $1$; moreover, we use the opposite sign
for the Chern-Simons terms in the definition of the R-R field strength.}, 
\begin{equation} 
%\gamma = c - \tau (b-1/2).  
\gamma =2\pi (c + \tau (b-1/2)).  
\label{3} 
\end{equation} 
%Normalizations have been chosen in such a way that $\gamma$ is 
%the difference between the coupling constants of the two groups. 
The supergravity equations of motion require an holomorphic $\gamma$. 
The linearized result~\cite{kn}, 
\begin{equation} 
\gamma (z)=\gamma^{(0)}+2i\left (\sum_{i=1}^{n_1}\log (z-z_i^{(1)})- \sum_{i=1}^{n_2}\log (z-z_i^{(2)})\right ). 
\label{5} 
\end{equation} 
combined with a black D3-brane ansatz  
\begin{eqnarray} 
ds^2&=&Z^{-1/2}dx_{\mu}dx^{\mu}+Z^{1/2}ds^2_K,\nonumber\\ 
F_5&=&dC_4+*dC_4,\qquad C_4={1\over Z}dx^0\wedge dx^1\wedge dx^2  
\wedge dx^3,
\label{6} 
\end{eqnarray} 
is a solution of type IIB equations of motion provided that 
\begin{equation} 
- \Box_K Z=\rho_{D3}(x) + {\rm const}~
|\partial\gamma(z)|^2\delta^{(4)}(x_6,x_7,x_8,x_9).
\label{7} 
\end{equation} 
Here $\rho(x)$ is an arbitrary density of physical D3-branes
\cite{pg,bertolini}. The general solution of this equation is 
\begin{equation} 
Z (x_T,z) =\sum_{i=1}^{n_1}{b^{(0)}\over (x_T^2+|z-z_i^{(1)}|^2)^2} 
+ \sum_{i=1}^{n_2}{1-b^{(0)}\over (x_T^2+|z-z_i^{(2)}|^2)^2}+{\rm const}\int d^2 w 
{|\partial\gamma (w)|^2\over (x_T^2+|z-w|^2)^2}. 
\label{8} 
\end{equation} 
 
The logarithmic behaviour in~(\ref{5}) reproduces 
the one-loop beta function of the ${\cal N}=2$ gauge theory~\cite{kn}. It 
is interpreted in the T-dual picture 
as the bending of the NS due to the D4-branes.  
  
The solution~(\ref{8}) 
presents various kinds of singularity.  
There is certainly an IR singularity 
and, more generally, we expect singularities at the positions of the constituent  
branes. 
In ${\cal N}=2$, the 
enhan\c con mechanism~\cite{jpp} is usually invoked:  
The branes should resolve the singularity 
by forming shells. The enhan\c con mechanism thus suggests a natural IR cut-off 
for the integral in eq.~(\ref{8}). Notice also that   
the warp factor  is singular for $x_T=0, z=w$, if 
$\partial\gamma (w)\ne 0$. This can be interpreted as the result of 
the break down of the supergravity approximation near the  orbifold 
fixed planes. Only if $\partial\gamma (w)$ has compact support,  
the solution is asymptotically ($z\gg 1$) well defined for all $x_T$.  
 
For supergravity and 
AdS/CFT purposes, it is  
better to move together large bunches of branes. We therefore 
consider the limit $N\gg M\gg 1$.  
As usual, we keep the t'Hooft parameters $x=Ng_s$ and $y=Mg_s$ fixed. 
Moreover, the choice of a $U(1)$  
invariant configuration helps in improving the  
UV behaviour of eq.~({\ref{8})\footnote{For example, it improves 
 the convergence of the spherical harmonic expansion  
of the integral in eq.~(\ref{8}).}. For these reasons, we choose the 
configuration analysed in Section 3~\cite{Pproc,a}: 
We take $M$ fractional and  $N$ physical branes
at $z=0$, and  $M$ anti-fractional branes in a rotational invariant  
configuration 
at $|z|=z_{\infty}$. 
 
The one-loop beta functions 
are, for generic $b^{(0)},c^{(0)}$, 
\begin{equation} 
%\tau_1=2iM\log{z\over\Lambda},\qquad\qquad  
%\Lambda=z_{\infty}e^{i(\tau b^{(0)} - c^{(0)})/2M}, 
2\pi\tau_1=2iM\log{z\over\Lambda},\qquad\qquad  
\Lambda=z_{\infty}e^{2\pi i(c^{(0)} + \tau b^{(0)})/2M}, 
\label{p3} 
\end{equation}  
and 
\begin{equation} 
2\pi\tau_2 = %2\pi(\tau-\tau_1)= 
-2iM\log{z\over\Lambda_2},\qquad\qquad  
\Lambda_2=z_{\infty}e^{2\pi i(c^{(0)} + \tau (b^{(0)}-1))/2M}, 
\label{p3bis} 
\end{equation}  
where the $\Lambda$'s are the dynamically generated scales. 
We explicitly took the limit $N\gg M\gg 1$ and considered  
$|z|<z_{\infty}$.  
Notice that $\Lambda_2$, which, according to this  
semi-classical reasoning, is the scale where the anti-fractional 
brane theory 
becomes strongly coupled, is above the cut-off and therefore it is not 
relevant to our analysis. Equations~(\ref{p3}),(\ref{p3bis}) reduce to 
eq.~(\ref{1-loop}) for $b^{(0)}=1/2$, $c^{(0)}=0$.  

An important point, worth to be stressed, is that the one-loop 
behaviour~(\ref{p3}),(\ref{p3bis}) is appropriate for not one but many
points in moduli space. The presence of 
physical branes at arbitrary points, for example,
does not change the result~(\ref{p3}),(\ref{p3bis}).  
 
In the limit $N\gg M\gg 1$, the solution in~(\ref{8})    
resembles the KS solution~\cite{ks} for ${\cal N}=1$. 
This is mainly due to the $\log$ in eq.~(\ref{5}) and may suggest 
new physics for ${\cal N}=2$, with  a cascade mechanism similar to that in 
\cite{ks}. Actually, eqs.~(\ref{2}),(\ref{p3}) 
indicate that the background passes many times through values 
where $b\in Z$ and non-perturbative phenomena become relevant 
\cite{Pproc,a}. Start, for simplicity, 
 at the cut-off with $b^{(0)}=1/2$, $c^{(0)}=0$. $b(z)$ decreases with 
$|z|$ and reaches the value $b(z)=0$ for $|z|=\Lambda$.  
A tensionless string phase in type IIB requires 
$b=c=0$, which (see eq.~(\ref{2})) is the same as $\tau_1=0$. 
As also noticed 
in~\cite{a}, this only selects $2M$ distinguished points on 
the circle $|z|=\Lambda$. 
They coincide with the branch-points for  
the curve discussed in Section 3 and represent an 
{\it enhan\c con}. The tension of a fractional brane probe is 
given by $\tau_1(z)$ and vanishes at the enhan\c con, suggesting 
that the probe cannot move below the scale $|z|=\Lambda$~\cite{Pproc,a}. 
An anti-fractional probe can  
instead move even below $|z|=\Lambda$, since its tension is given by 
$\tau_2(z)$, which is non-vanishing. This perfectly agrees with our 
discussion based on the SW curve in Section 3. 

The basic puzzle about the supergravity solution regards the scale
where corrections actually start modifying it.
Instantonic and higher derivative corrections are particularly
complicated in these systems because of the presence of many
other effects, for example tensionless string phases.
 
The crucial question is what happens below $|z|=\Lambda$. 
An extrapolation of the log behaviour would suggest not one, but many 
enhan\c cons, whose physics needs to be explained, for example via 
a Seiberg duality for ${\cal N}=2$~\cite{Pproc}.   
The phenomena discussed in~\cite{Pproc} could in principle apply to some  
configurations, but it is difficult to make more precise statements.
We focused on the origin of moduli space.  
If we take the attitude that the string resolution mechanism  
in type IIB should be the same as the one suggested by the SW curve, 
discussed in Section 3,  
we would conclude that there is only one enhan\c con at $\Lambda$.
This is the point where tensionless strings and non-perturbative
phenomena may become relevant and modify the semi-classical background. 
From the SW curve, we could expect that the 
supergravity fields for $|z|<\Lambda$ are frozen at the value 
they attained at the enhan\c con. The picture would be 
similar to the first example of enhan\c con discussed in~\cite{jpp}. 
The presence of physical branes should not modify 
the physics too much, at least when they sit at the origin 
of moduli space. In contrast with ${\cal N}=1$, here there is a moduli 
space both for physical and fractional branes. 
We could have constructed our system by first sending $N$ physical branes 
to the origin and only then trying to send in the $M$ fractional ones. 
A reasonable expectation is that they form a spherical shell very 
similar to that for pure $SU(M)$ theory.  

A standard observation 
in favour of the existence of many enhan\c con made
in~\cite{Pproc,a} is that composite probes with 
both physical and fractional charges should be able to
move below $\Lambda$ and stop at a successive enhan\c con,
the point where their tension vanishes. Take for
example a probe with charges $(3/2,1/2)$ in the orbifold 
background ($b=1/2$). We can realize it by considering a bound state of
a physical and
a fractional brane. Notice that this system has many moduli: 
A fine tuning is required  to move it as a single composite object.
The log behavior suggests that its tension is finite 
at $\Lambda$ and vanishes at a scale $\bar z<\Lambda$, which may define
the second enhan\c con.
However, as noticed in~\cite{Pproc},
the system stops being BPS at $\Lambda$. We can interpret this
phenomenon as the fact that the fractional brane 
is obliged to stop at $\Lambda$,
while the remaining physical brane is free to move below and reach
the origin. This is consistent with our previous discussion. 
   
Let us conclude this Section by discussing the decoupling limit  
$z_{\infty}\rightarrow\infty$.  
If $N=0$ we obtain the pure $SU(M)$ 
theory, which, as a quantum field theory, is defined at all scales. 
From eq.~(\ref{p3}), 
we see that the correct limit for removing the cut-off is 
$z_{\infty}\rightarrow\infty, Mg_s\rightarrow 0$. 
Since, for $N=0$, the t'Hooft parameter  $Mg_s$ has to be large for 
supergravity to be valid, this limit necessarily involves a string theory 
description. As usual, this was expected since the theory is  
asymptotically free in the UV. 
 
For $N\ne 0$ we have a second option~\cite{Pproc,a}.  
We can take $x=Ng_s$ finite and 
large, so that all curvatures in the solution are small, while 
keeping $y=Mg_s$ fixed.  
In the decoupling limit  
$z_{\infty}\rightarrow\infty, y\rightarrow 0$,   
we obtain the theory $SU(N+M)\times SU(N)$. 
The scale for the asymptotically free factor $SU(N+M)$ is $\Lambda$. 
However, the theory is 
not well-defined since one of the gauge factors is not asymptotically  
free. The point $\Lambda_2$, which signals non-perturbative effects for 
the $SU(N)$ gauge group, is now sent all the way to infinity. 
We expect that the one-loop 
behaviour is accurate for $z>\Lambda$, but fails in the UV, where 
new degrees of freedom are required for the QFT. 
This could be signaled, on the supergravity side, by D1-instanton
effects.  

In this paper, we mainly considered the origin of moduli space of
the ${\cal N}=2$ theory. In such a point, the presence of physical
branes does not affect most of the reasonings.
Other points are certainly interesting.
It was suggested in~\cite{a} that a log solution with
multiple enhan\c cons is the description of a point of the
moduli space with distributed physical branes. They should explain
the logarithmically varying five-form flux. The large $N$ limit 
considered in~\cite{a} seems to be different from that considered
in this paper.
 
\vskip .2in \noindent \textbf{Acknowledgments}\vskip .1in \noindent   
This work has been supported by the Swiss National Science
Foundation, by the European Union under RTN contract
HPRN-CT-2000-00131. A.~Z. is partially supported by INFN and MURST. 
R.~R. would like to thank the authors of~\cite{bertolini} and Marco
Bill\`o for continous exchange of information and for many interesting
discussions. 

\section*{Appendix: Comments on the AdS/CFT interpretation} 
In this Appendix we give a small
dictionary for the AdS/CFT correspondence applied to our system,
and discuss few subtleties.  
For the conformal case $SU(n)\times SU(n)$,
the supergravity solution~(\ref{6}) is well  defined and   
becomes AdS$_5\times$S$_5/Z_2$.
%Without loss of generality,  
%we consider the conformal theory 
%$n_1=n_2=n$.  
%All other theories can be obtained by considering 
%suitable limits in the moduli space. For $n_1=n_2$ the 
%supergravity solution~(\ref{6}) is well  defined and  
%approaches AdS$_5\times$S$_5/Z_2$. 
One can then use  AdS/CFT 
to describe the conformal field theory corresponding to the 
origin of moduli space, where Higgs and Coulomb branch meet. 
We could also use the standard rules of AdS/CFT
to study the Coulomb branch. 

%The mapping between supergravity fields and CFT operators  
%is known in details. 
We first discuss the $U(1)$'s behaviour  
in the various pictures used in the paper. 
The system of $n$  
fractional and anti-fractional D3 branes stuck at a $Z_2$ 
orbifold point in type IIB has a moduli 
space which is isomorphic to the Coulomb branch 
 of an ${\cal N}=2$, $U(n)\times U(n)$ gauge theory.  
The diagonal $U(1)$ factor 
is decoupled and corresponds to the center of mass motion of the system. 
$U(1)$'s factors may disappear in different representations and 
limits. The corresponding moduli are frozen and may reappear 
as mass parameters in the gauge theory\footnote{This happens, 
as usual, because of the ${\cal N}=2$ standard coupling  
$A\Phi B$ between the adjoint 
fields $\Phi$ (in ${\cal N}=1$ notations) and the hypermultiplets $(A,B)$.}. 
In the picture with D4 and NS-branes, $U(1)$ factors 
are usually frozen~\cite{witten}. In elliptic models,  
the diagonal $U(1)$ is present and decoupled, while the second $U(1)$  
is frozen because $m=\sum z_i^{(1)}-\sum z_i^{(2)}$ is not 
normalizable~\cite{witten}.  
The centers of mass of the two sets of D4-branes can be nevertheless at 
different points in $x_4+ix_5$. $m$ 
has to be interpreted not as a modulus 
but as a mass term   
for the hypermultiplets.  
%The other independent mass term can be 
%introduced with a twist in the M theory lifting of the configuration 
%\cite{witten}, but we will not discuss this term in this paper.  
In the AdS/CFT description of this system, all the $U(1)$ factors are, 
as usual, absent and the gauge group is $SU(n)\times SU(n)$. 
Supergravity solutions with non-zero  $m$ 
are interpreted as mass deformations of the $SU(n)\times SU(n)$ CFT. 
The  $Z_{2M}$ rotationally invariant 
configuration for anti-fractional branes in Sections 3 and 5 
corresponds to choice of a point in the Coulomb branch. A configuration
with all anti-fractional branes at a specific point $z_{\infty}$
would correspond instead to a deformation of the CFT with a mass term
combined with a choice of vacuum.
 
We now discuss the mapping between CFT operators and supergravity  
fields. 
For the untwisted fields, it follows from a $Z_2$  
projection of the parent AdS$_5\times$S$^5$ theory. For the twisted 
fields, the mapping was explicitly worked out in~\cite{g}. 
The order parameters for the Coulomb branch are associated with  
the operators 
\begin{eqnarray} 
O_k&=&{\mbox Tr}(b^{(0)}\phi^k_{(1)}+(1-b^{(0)})\phi^k_{(2)}),\,\,  
k=2,3,...\nonumber\\ 
T_k&=&{\mbox Tr}(\phi^k_{(1)}- \phi^k_{(2)}),\,\, k=2,3,... 
\label{ads1}\end{eqnarray} 
$O_k$ couple to the untwisted fields, specifically to the 
spherical harmonics of the metric. 
$T_k$ couple to the harmonics of the twisted fields 
$\gamma(z)=\sum_n \gamma_n z^n,n\in Z$. 
As discussed in Section 2, the zero mode 
$n=0$ corresponds to the dimension four operator ${\mbox Tr}(F^2_{(1)}- 
F^2_{(2)})$, dual to the coupling $\tau_1-\tau_2$. 
The other harmonics  correspond to the 
operators $T_k$ 
and $W_i= {\mbox Tr}(F^2\phi^k_{(1)}-F^2\phi^k_{(2)})$, $k=1,2,...$. 
The only subtlety is that the $n=1$ mode is associated with
the dimension three operator $\int d^2\theta {\mbox Tr}(B_1A_1-B_2A_2)$.
$T_1$ is indeed identically zero.
The absence of non-trivial dimension 1 operators in any conformal theory
and a quick check to the mass spectrum found in~\cite{g} confirm this
identification, which is consequence of the $U(1)$'s disappearing.
The mapping operators/fields and the explicit coefficients in the 
definition~(\ref{ads1}) follow from the Born-Infeld 
action for wrapped D5-branes. 
 
According to the  
standard AdS/CFT interpretation of supergravity solutions approaching   
AdS in the UV~\cite{bala},   
the asymptotic behaviour of a supergravity field (with scaling dimension 
$\Delta$) $\psi_O\sim z^{-\Delta}$ corresponds to 
 a VEV for the corresponding operator $O$, while $\psi_O\sim z^{\Delta-4}$ indicates that the CFT is deformed with $O$.  
The UV expansion ($z\gg 1$) of  equation~(\ref{5}) indeed suggests that  
the operators $T_k$ have a VEV 
$\sum_i (z_i^{k (1)}-z_i^{k (2)})$, as expected for a generic point of 
the Coulomb branch. The leading term 
$\gamma\sim (z_i^{(1)}-z_i^{(2)})/z=m/z$ is appropriately interpreted 
as a deformation with a dimension three operator, rather than 
a VEV for a dimension one operator. As discussed above, this is
the mass term  
$\int d^2\theta {\mbox Tr}(B_1A_1-B_2A_2)$. 
A mass for hypermultiplets also requires a scalar mass 
term. We can find it in one of the harmonic 
for the metric, $Y_2(\hat x)=\sum_{a=6}^9\hat x_a^2-2\hat x_4^2-2\hat
x_5^2, \sum_{I=4}^6\hat x^2=1$.   
For a correct AdS/CFT interpretation, 
such deformation with a dimension two operator should  
affect the expansion of the warp factor $Z\sim (1+m^2Y_2(\hat x) 
\log (r)/r^2)/r^4$ 
for large $r$. Qualitatively, this is easily extracted from the integral in eq.~(\ref{8}) 
for $\gamma\sim m/z$. 

Finally, we could extract information about the VEV's of $O_k$ from
the large $r$ expansion of eq.~(\ref{8}).  The expansion of the first
two terms in the right hand side is straightforward~\cite{bala}. The
third term requires particular care because of the integration over
the entire plane $z$ and the many types of divergences. Some
assumptions and {\it renormalizations} are in general needed.

\end{document}